\documentclass[10pt,letterpaper,twocolumn]{article}
\usepackage[english]{babel}
\usepackage{graphicx}

\newcommand{\alt}{\mathbin{\lower 3pt\hbox
   {$\rlap{\raise 5pt\hbox{$\char'074$}}\mathchar"7218$}}}
\newcommand{\agt}{\mathbin{\lower 3pt\hbox
   {$\rlap{\raise 5pt\hbox{$\char'076$}}\mathchar"7218$}}}

\begin{document}

\setcounter{footnote}{0}
\setcounter{equation}{0}
\setcounter{figure}{0}
\setcounter{table}{0}

\title{\large\bf Conductance distribution in the magnetic
field}

\author{\small I. M. Suslov  \\
\small P.L.Kapitza Institute for Physical Problems,  \\
\small 119334 Moscow, Russia  \\
\small E-mail: suslov@kapitza.ras.ru\\
{}\\
\parbox{120mm}{\footnotesize \,\,Using a modification of the
Shapiro scaling approach, we derive the distribution
of conductance in the magnetic field applicable in the vicinity of
the Anderson transition. This distribution is described by
the same equations as in the absence of a field.
Variation of the magnetic field does not lead to any
qualitative effects in the conductance distribution
and only changes its quantitative characteristics, moving
a position of the  system  in the three-parameter space.
In contrast to the original Shapiro approach, the evolution
equation for quasi-1D systems is established from the generalized DMPK
equation, and not by a simple analogy with one-dimensional
systems; as a result, the whole approach became more rigorous
and accurate.  } }

\date{}
\maketitle

\textwidth 6.4 in
\textheight 8.5 in

\setcounter{footnote}{0}
\setcounter{equation}{0}
\setcounter{figure}{0}
\setcounter{table}{0}

\begin{center}
{\bf 1. Introduction}
\end{center}

It is well-known that conductance of a disordered system is a
strongly fluctuating quantity: its root-mean-square deviation
in the metallic phase does not depend on the system size \cite{1,2}
and becomes comparable with its mean in the vicinity of
the Anderson transition; as a result the problem of its
distribution $W(g)$ arises. Here and below $g=hG/e^2$ is a
dimensionless conductance, which is determined by conductance
$G$ of a sample in quantum units  $e^2/h$.

In the recent paper \cite{3}, using a modification of the Shapiro
approach \cite{4,5},
we have introduced the two-parameter family
of conductance distributions $W(g)$, which is in
one-to-one correspondence with conductance distributions
of quasi-1D systems of size $L^{d-1}\times L_z$ ($d$ is a
dimension of space), characterizing by parameters  $L/\xi$
and $L_z/L$ ($\xi$ is the correlation length).
Investigation of this family allowed to describe all essential
features of the distribution $W(g)$, established in numerical
experiments, and reproduce the results for its cumulants
derived in the sigma-model formalism \cite{101}.
The approach of \cite{3} is
based on the evolution equation for the distribution
$P(\rho)$ of dimensionless Landauer resistances \cite{7}
($\rho=1/g$) for  one-dimensional systems, which are
arranged by a certain scheme to compose the
$d$-dimensional system \cite{4,5}.
At first glance, generalization of
these results \cite{3} for the case of
a non-zero magnetic field
presents no problem: one should only use the  1D
evolution equation without assumption of time-reversal
invariance. However, realization of this scheme (Sec.2)
meets two difficulties:  (a) disagreement in the
number of essential parameters, and (b) incorrect
estimation of the critical behavior in  $2+\epsilon$
dimensions. Analysis of these contradictions leads to conclusion
(Sec.3) that they are related with a qualitative difference
between quasi-1D and strictly
one-dimensional systems in  presence of the
magnetic field. Replacement of the 1D
evolution
equation by the generalized Dorokhov--Mello--Pereyra--Kumar
(DMPK) equation  \cite{8} eliminates the indicated
problems (Sec.4). The evolution equation for $P(\rho)$
derived from the generalized DMPK equation (Sec.5) has the same
structure as in the 1D case but with variable coefficients.
The latter explains the origin of indicated difficulties, but
becomes inessential after transition to the $d$-dimensional
case (Sec.6). As a result, the conductance distribution in the
magnetic field is described by the same equations as in the
absence of a field; however, one of irrelevant parameters
becomes relevant. Therefore, variation  of the magnetic field
does not lead to any qualitative effects in the conductance
distribution and only changes its quantitative characteristics,
moving a position of  the system in the three-parameter
space.  These conclusions are in accordance with numerical
experiments \cite{9,10}.

\begin{center}
{\bf 2. The simplest scheme}
\end{center}

The approach used in the paper \cite{3} is based on the
large-scale constructions by Shapiro \cite{4,5},
analogous to the Migdal--Kadanov transformation in the
usual phase transitions theory \cite{11,12}. Firstly, $b$
cubical blocks of size $L$ are arranged successively to form a
quasi-one-dimensional system (Fig.1), and then a parallel
connection of  $b^{d-1}$ quasi-1D chains composes the
$d$-dimensional cube of size $bL$.  For simplicity, the quasi-1D
systems are supposed to be isolated by dielectric inter-layers
(Fig.1);  however, in the large $L$ limit the concentration of
the auxiliary dielectric phase tends to zero, and the Shapiro
scheme looks well-grounded.  Only the latter limit will be of
interest for us (see a detailed discussion in [3]).

\begin{figure}
\centerline{\includegraphics[width=2.5 in]{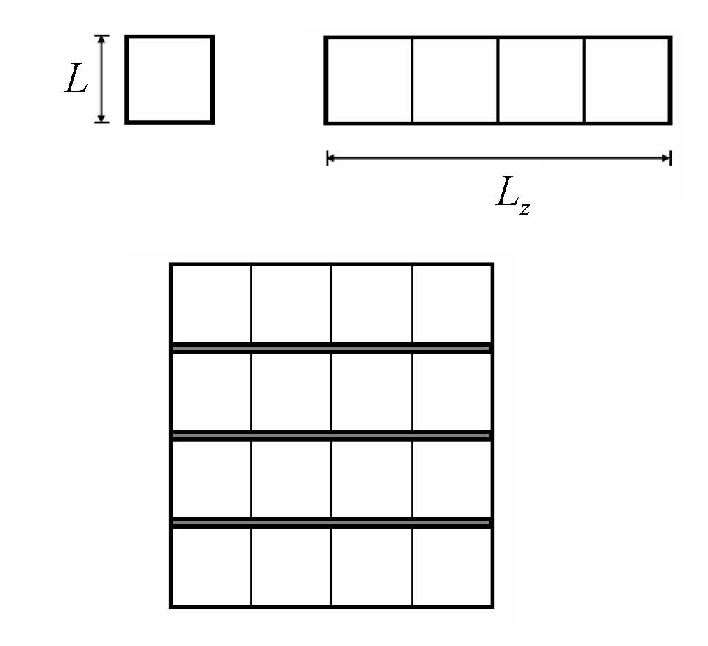}}
\caption{ Large-scale constructions used in the
Shapiro scheme: the cubical blocks of size $L$ are
arranged into quasi-1D systems of length $L_z$, whose
parallel connection composes the  $d$-dimensional cube of
the larger size.  } \label{fig1}
\end{figure}

According to the one-parameter scaling hypothesis \cite{13}, all
properties of the cubic system of size $L$ are completely
determined by the ratio $L/\xi$. The properties of the quasi-1D
system, composed of cubical blocks, are specified by the
properties of the single block ($L/\xi$) and  a number of cubes
($L_z/L$), so  for conductance
$$
g=F\left(\,\frac{L}{\xi}\,, \,\frac{L_z}{L}\,\right) \,.
\eqno(1)
$$
Setting $L=a$  ($a$ is the atomic spacing), one
comes to conclusion, that the conductance distribution
$W(g)$ of the quasi-1D system corresponds to a
certain conductance distribution of the strictly
one-dimensional system\,\footnote{\,We shall
return later to validity of this statement. For the moment it
suffices to note
that this statement is rigorous in
the framework of the orthodox scaling of the paper \cite{13}.}.
This conclusion agrees with the fact that the well-known
evolution equation for  1D systems
\cite{4,14,15,16,17}
$$
\frac{\partial P(\rho)}{\partial L} =\alpha
\frac{\partial }{\partial \rho}
\left[\,\rho(\rho+1) \frac{\partial P(\rho)}{\partial \rho}
\,\right]
\eqno(2)
$$
allows two-parameter generalization
$$
\frac{\partial P(\rho)}{\partial L} = \tilde\alpha\,
\frac{\partial }{\partial \rho}
\left[\,-\gamma(2\rho+1) P(\rho) + \rho(\rho+1)
\frac{\partial }{\partial \rho} P(\rho) \,\right] \,,
\eqno(3)
$$
so parameters $L/\xi$  and  $L_z/L$ of equation (1) are
in one-to-one correspondence with parameters  $\tilde\alpha L$
and $\gamma$ specifying the solution of (3). We do not try to
establish a character of this correspondence for any specific
situations but investigate all family of distributions in whole.
The properties of a quasi-1D system are described by
equation (3) and can be considered as known in principle, so
transition to the $d$-dimensional case presents no problem: it is
sufficient to come from $P(\rho)$ to  $W(g)$ and find the
distribution of a sum of $b^{d-1}$ independent random quantities
with the same distribution  $W(g)$. This procedure can be
realized in the differential form \cite{3,4,5} and leads to the
evolution equation describing the $d$-dimensional system
(Sec.6).

\begin{figure}
\centerline{\includegraphics[width=2.5 in]{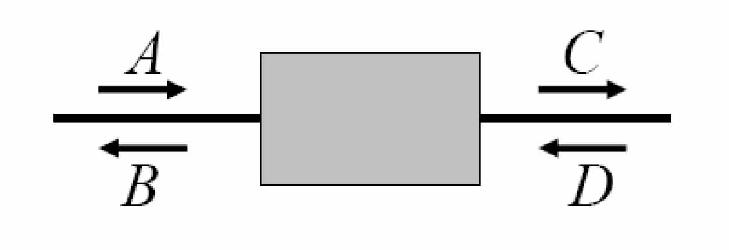}}
\caption{In one-dimensional systems, the transfer matrix $\hat
T$ relates amplitudes of plane waves on the left and on the right
of a scatterer.  } \label{fig2} \end{figure}

In description of one-dimensional systems, it is convenient
to consider each scatterer as  \mbox{"a black box"},
characterizing by a transfer matrix $\hat T$, relating
amplitudes of the plane waves on the left  ($Ae^{ikx}+Be^{-ikx}$)
and on the right ($Ce^{ikx}+De^{-ikx}$) of
the scatterer  (Fig.2):
$$
\left ( \begin{array}{cc} A \\ B \end{array} \right)\,
= \hat T \left ( \begin{array}{cc} C \\ D \end{array}
\right) \,.
\eqno(4)
$$
A successive arrangement of scatterers corresponds
to multiplication of transfer matrices. In the presence of
time-reversal invariance, the transfer matrix allows a
parametrization  \cite{18}
$$
\hat T= \left ( \begin{array}{cc} \sqrt{\rho\!+\!1}\, e^{-i\varphi} &
-\sqrt{\rho} \,e^{-i\theta}
\\ -\sqrt{\rho}\, e^{i\theta} & \sqrt{\rho\!+\!1}\,
e^{i\varphi} \end{array} \right)\,,
\eqno(5)
$$
so one should consider the mutual distribution function
$P\left(\rho, \varphi, \theta\right)$ for
the parameters entering (5). In the product of $n$
transfer matrices the distribution of phases $\varphi$ and
$\theta$ is usually stabilized for large $n$, so
$$
P_n\left(\rho, \varphi, \theta\right)=
P_n\left(\rho\right) P\left(\varphi, \theta\right)\,.
\eqno(6)
$$
If distribution of phases is uniform
($P\left(\varphi, \theta\right)=const$), then equation
(2) is valid, while in the general case one obtains Eq.3
with parameters
$$
\gamma=\frac{1-2A_0}{2A_0},
\quad \tilde\alpha =2\alpha A_0,
\quad A_0= \left\langle \sin^2(\varphi-\theta) \right\rangle
.
\eqno(7)
$$

From the above discussion one can see a simple way to
generalize the described procedure to the case of the non-zero
magnetic field. It is sufficient to replace the transfer matrix
(4) by the more general expression
$$
\hat T= \left ( \begin{array}{cc} \sqrt{\rho\!+\!1}\, e^{-i\varphi} &
-\sqrt{\rho} \,e^{-i\theta+i\zeta}
\\ -\sqrt{\rho}\, e^{i\theta} & \sqrt{\rho\!+\!1}\,
e^{i\varphi+i\zeta} \end{array} \right)\,,
\eqno(8)
$$
which is valid without assumption of the time-reversal
invariance\footnote{\,For definiteness, we keep in mind the
case of the external magnetic field, while the following analysis
is equally applicable for the case of the magnetic
impurities.}, and repeat
derivation of the evolution equation (see Appendix 1); as
a result, we come to the same equation  (3) with parameters
$$
\gamma=\frac{1-2A_0}{2A_0},
\quad \tilde\alpha =2\alpha A_0\,,
\quad A_0= \left\langle \sin^2(\varphi-\theta+\zeta)
\right\rangle.
\eqno(9)
$$
One can see that a presence of the magnetic field leads to
the only effect that the phase  $\zeta$, being strictly equal to
zero in the absence of a field, acquires a certain stationary
distribution; this changes coefficients of Eq.3 but does not
change its structure.
Using results of  \cite{3},
we come to conclusion that the conductance distribution in the
$d$-dimensional case is described by the same equations, as in
the absence of a field.

However, on a closer examination
the described scheme meets
two difficulties. Firstly, in the presence of the magnetic
field, Eq.1 is replaced by
$$
g=F_H\left(\,\frac{L}{\xi}\,, \,\frac{L_z}{L}\,,
\,\frac{L}{l_H}\,\right) \,,
\eqno(10)
$$
where $l_H=(c\hbar/2eH)^{1/2}$ is the magnetic length. Setting
$L=a$, we see that
in description of one-dimensional systems one
should have three essential parameters, and not two, as in Eq.3.


Secondly, there is a problem with estimation of the critical
behavior of the correlation length $\xi$, given in Sec.5.1
of \cite{3} and initially suggested by Shapiro \cite{5}.
Multiplying Eq.2 by $\rho$ and integrating, one has a closed
equation for the average value  $\left\langle\rho\right\rangle$,
whose solution
$$
\left\langle\rho_L\right\rangle={\textstyle\frac{1}{2}}
\left(e^{2\alpha L}-1 \right) \eqno(11)
$$
can be rewritten in the form of the scale transformation for
1D systems
$$
\left\langle\rho_{bL}\right\rangle
={\textstyle\frac{1}{2}}
\left[\left(1+2\left\langle\rho_{L}\right\rangle
\right)^b -1\right] \,.
\eqno(12)
$$
For the parallel connection of $b^{d-1}$ one-dimensional chains,
resistance is diminished by a factor $b^{d-1}$, and one has a
scale transformation for the $d$-dimensional system:
$$
\left\langle\rho_{bL}\right\rangle
={\textstyle\frac{1}{2}} b^{-(d-1)}
\left[\left(1+2\left\langle\rho_{L}\right\rangle
\right)^b -1\right]\,.
\eqno(13)
$$
If Eq.3 is used instead of (2), then dependence on  $\gamma$
disappears and results (11--13) remain unchanged. In
fact, for the parallel connection of chains the average
conductances should be added, i.e.
 $\left\langle g_{bL}\right\rangle =
b^{d-1} \left\langle g_{L}\right\rangle$ instead of the
previously used relation $\left\langle
\rho_{bL}\right\rangle= b^{-(d-1)} \left\langle
\rho_{L}\right\rangle$. As a result,
the scale transformation (13) is valid only for
$d=2+\epsilon$, when the distribution $P(\rho)$ is narrow and two
indicated relations approximately coincide; in this case
one obtains  the correct result $\nu=1/\epsilon$
for the critical exponent of the correlation length.  Since in
the presence of the magnetic field equation (3) remains
unchanged, then the result $\nu=1/\epsilon$ persists as before;
but now it is incorrect, because one should have $\nu=1/2\epsilon$
\cite{19,22}.

 We meet  a strange situation: the use of the Shapiro
 scheme gives excellent results in the absence of a magnetic
 field \cite{3}, but  leads to evident
 contradictions when the magnetic field is present.

\begin{center}
{\bf 3. Analysis of the situation}
\end{center}


To analyze the situation, we  use the fact that
the critical behavior for  $d=2+\epsilon$ is related with effects
of weak localization and allows a simple physical interpretation.

\begin{figure}
\centerline{\includegraphics[width=2.5 in]{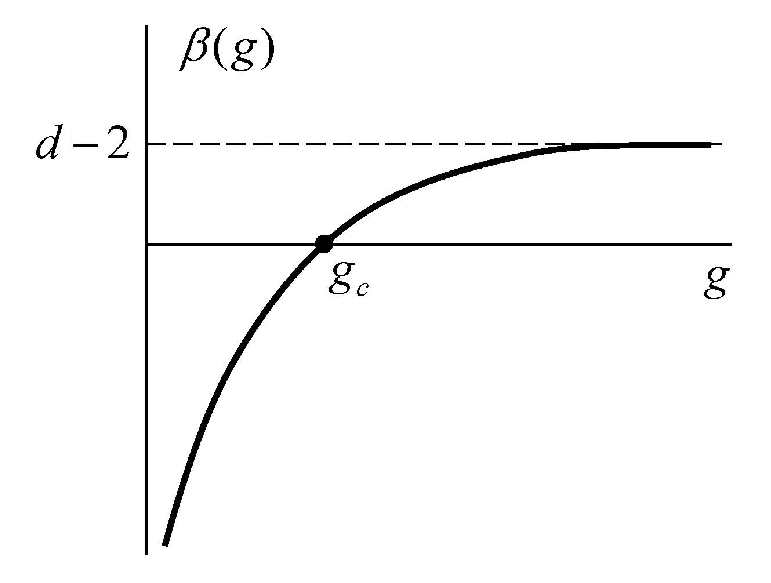}}
\caption{A typical behavior of the function $\beta(g)$
in equation  (14).  }
\label{fig3}
\end{figure}

According to the hypothesis of one-parameter scaling \cite{13},
the typical value of conductance  $g$ obeys the
renormalization group equation
$$
\frac{d\ln g}{d \ln L}=\beta(g) \,,
\eqno(14)
$$
where  $\beta(g)=d-2$ for $g\to\infty$ and $\beta(g)=\ln g$
for $g\to 0$. For $d>2$ the function  $\beta(g)$ has a
root $g_c$ (Fig.3), which corresponds to the Anderson transition
point; the critical exponent $\nu$ is determined by the
derivative of $\beta(g)$ at $g_c$:
$$
\frac{1}{\nu}=g_c \beta'(g_c)\,.
\eqno(15)
$$
For $d=2+\epsilon$ the critical point is located in
the region of large $g$, where expansion in $1/g$ is
possible
$$
\beta(g)=\epsilon+\frac{A_1}{g}+\frac{A_2}{g^2}+\ldots
\eqno(16)
$$
It is easy to verify that retaining two first terms in (16)
we have  for $A_1<0$
$$
\nu=1/\epsilon\qquad \mbox{\rm (independently of $A_1$)}\,,
\eqno(17)
$$
while for $A_1=0$ and $A_2<0$
$$
\nu=1/2\epsilon\qquad \mbox{\rm (independently of $A_2$)} \,,
\eqno(18)
$$
i.e. in the main  $\epsilon$-approximation the
critical behavior is determined by the structure of
expansion  (16), and not by specific values of coefficients.

For $d=2$ integration of (14) with the initial condition
$g=g_0$ at $L=a$  gives in the large  $g$ region
$$
g=g_0+A_1\ln(L/a)  \,,
\eqno(19)
$$
i.e. well-known logarithmic correction of the weak localization
theory. Its existence can be controlled by the diagrammatic
approach \cite{20}, confirming finiteness and
negativeness of $A_1$. In the presence of the magnetic field, the
logarithmic divergency for $L\to\infty$ is cut off at the magnetic
length $l_H$, and absence of the contribution $\sim \ln L$
signifies disappearance of the coefficient $A_1$ and validity of
the result  (18).


According to the qualitative picture of weak localization
\cite{22,21}, the main quantum correction to a classical
diffusion is related with self-intersection of
trajectories, when a possibility to pass the closed loop in
two opposite directions makes the quantum interference to be
inevitable. If a diffusion trajectory is represented as a tube
with the thickness of  order  the de Broglie wavelength
$\lambda$, then the probability of
self-intersection is determined by a ratio of the volume
$v_F\lambda^{d-1} dt$, sweeping by the trajectory
during time
$dt$, to the volume $(Dt)^{d/2}$ of the region where the
trajectory is localized at the instant of time  $t$
($v_F$ is the Fermi velocity, and
$D$ is a diffusion constant).  The main quantum correction
$\Delta g$ to the classical conductance $g_0$ is determined by
the total probability of self-intersection given by integration
over $t$
$$
-\frac{\Delta g}{g_0}\sim
\int v_F\lambda^{d-1} (Dt)^{-d/2} dt\,,
\eqno(20)
$$
and has logarithmic divergency at the upper limit for $d=2$
(the lower limit of integration is given by the elastic
mean free time $\tau$). This divergency is cut off at the
scale $\tau_L\sim L^2/D$ and leads to the logarithmic $L$
dependence.

In the presence of the magnetic field, the amplitudes of
bypassing the closed loop in two opposite directions acquire
the phase difference $\Delta\varphi$, which becomes
comparable with $2\pi$ for the loop area  of order
 $l_H^2$. For loops of the greater size the quantum correction is
destroyed, and logarithmic divergency for $L\to\infty$ is cut off
at the scale $l_H$. As a result, we have the complete physical
picture explaining the effects of weak localization and the
 critical behavior for  $d=2+\epsilon$.

For quasi-1D systems the closed loops are strongly oblong
(Fig.4) and their area tends to zero in the limit of strictly
one-dimensional systems. In the latter case,  the notion of
self-intersecting trajectories still survives, but  the effect
of destroying the quantum correction by the magnetic field
disappears completely. As a result, we come to inevitable
conclusion: the difference between quasi-1D and strictly 1D
systems is not very essential in  absence of a magnetic
field, but acquires a qualitative character in  presence of
the field.

\begin{figure}
\centerline{\includegraphics[width=2.8 in]{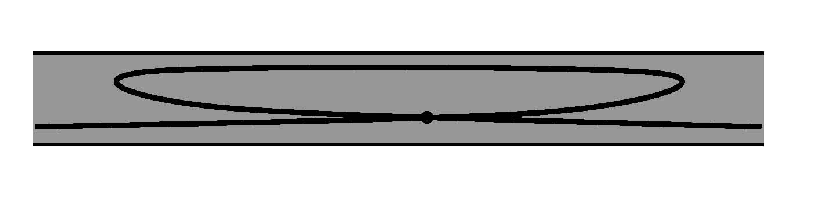}}
\caption{In quasi-1D systems, the closed loops are strongly
oblong, and their area tends to zero in the limit of strictly
one-dimensional systems. } \label{fig4}
\end{figure}

Let return to the possibility to set $L=a$ in Eq.1. In fact,
universal relations of such kind arise at the large length scales,
while for scales of order $a$ have a certain transient
behavior. If the function  $F$ in Eq.1 is assumed unchanged,
we can set only  $L\sim a$, but not $L=a$. This difference was
inessential in absence of a magnetic field, but becomes
the matter of principle in  presence of the field. In the
latter case, the function $F$ in Eq.1 should be related not to
equation (3), but with the evolution equation valid for quasi-1D
systems.

\begin{center}
{\bf 4. Generalized DMPK equation}
\end{center}

The counterpart of equation (2) in quasi-1D systems is given by
the Dorokhov--Mello--Pereyra--Kumar equation \cite{23}--\cite{27},
describing evolution of the diagonal elements of the
many-channel transfer matrix.

\begin{figure}
\centerline{\includegraphics[width=2.8 in]{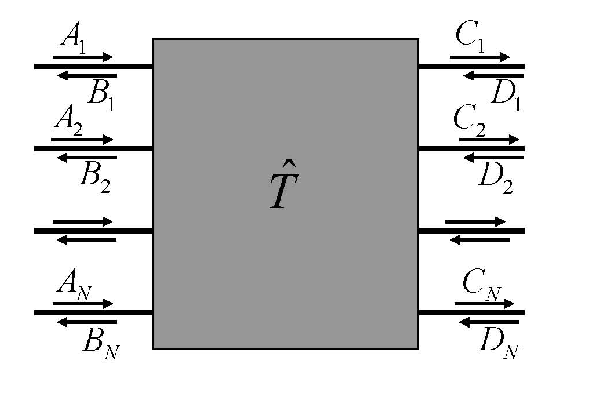}}
\caption{The many-channel transfer matrix $\hat T$ relates
the amplitudes of plane waves on the left ($A_n$, $B_n$)
and on the right ($C_n$, $D_n$) of a scatterer.  }
\label{fig5} \end{figure}

Considering the system as a set of $N$ coupled
one-dimensional chains, we can  treat it as an effective
scatterer and describe by the transfer matrix  $\hat T$,
relating the amplitudes of waves on the left
($A_n e^{ikx}+B_n e^{-ikx}$ in the $n$-th channel) and on the
right ($C_n e^{ikx}+D_n e^{-ikx}$) of the scatterer  (Fig.5).
Interpreting amplitudes $A_n$ as the components of the vector
$A$, and analogously for other amplitudes, we can write the
vector analog of relation (4)
$$
\left
( \begin{array}{cc} A \\ B \end{array} \right)\, = \hat T \left (
\begin{array}{cc} C \\ D \end{array} \right) \, = \left (
\begin{array}{cc} T_{11} & T_{12} \\ T_{21} & T_{22} \end{array}
\right)\, \left ( \begin{array}{cc} C \\ D \end{array} \right)
\,,
\eqno(21)
$$
where the transfer matrix $\hat T$ consists of four blocks of size
$N\times N$. It allows a parametrization \cite{25,28}
$$
\hat T= \left ( \begin{array}{cc}
 u_1 & 0 \\ 0 & v_1 \end{array} \right)\, \left (
\begin{array}{cc} \sqrt{1+\lambda} & \sqrt{\lambda} \\
\sqrt{\lambda} & \sqrt{1+\lambda} \end{array} \right)\,
	 \left ( \begin{array}{cc} u & 0 \\
         0 & v \end{array} \right)\,,
\eqno(22)
$$
where $u$, $v$, $u_1$, $v_1$ are unitary matrices, and
$\lambda$ is a diagonal matrix with the positive elements
$\lambda_i$, which in particular determine the conductance
$$
g_{ES}=\sum\limits_{i} \frac{1}{1+\lambda_i}\,
\eqno(23)
$$
(in the Economou--Soukoulis definition  \cite{29,30}). The DMPK
equation describes evolution of their mutual distribution
function  $P\left(\lambda_1, \lambda_2, \ldots, \lambda_N
\right) \equiv P\{\lambda\}$ with increasing  the length
$L$ of the system
$$
\frac{\partial P\{\lambda\}}{\partial L} = \alpha\,
\sum\limits_{i} \frac{\partial }{\partial \lambda_i}
\left[\,\lambda_i (1+\lambda_i)\, J\{\lambda\}
\frac{\partial }{\partial \lambda_i}\,
\frac{P\{\lambda\}}{J\{\lambda\}} \,\right]\,,
\eqno(24)
$$
$$
J\{\lambda\} = \prod\limits_{i<j} |\lambda_i-\lambda_j|^\beta
\,,
$$
where $\beta=1$ for the orthogonal ensemble (usual systems
with a random potential), $\beta=2$ for the unitary ensemble
(systems in the strong magnetic field), $\beta=4$ for the
symplectic ensemble (systems with the strong spin-orbit
interaction); the parameter  $\alpha=1/\xi_{1D}$ has a sense of
the inverse correlation length of the quasi-1D system. Equation
(24) is obtained from the maximum entropy principle (i.e. in
assumption of the maximal randomness compatible with the
symmetry restrictions) and ideologically close to the random
matrix theory by Wigner and Dyson \cite{31}.

In a strictly one-dimensional system one has $J\{\lambda\}=1$,
and $\lambda$ coincides with the Landauer resistance  $\rho$,
so (24) reduces to (2).\,\footnote{\,In this case it does not
contain the parameter  $\beta$, distinguishing the
orthogonal and unitary ensembles. It agrees with the previously
made conclusion that a strictly 1D system does not "feel"
the magnetic field. } This is quite natural, since (24) and
(2) are based on analogous assumptions (compare  \cite{17} and
\cite{24}). In the paper  \cite{8} we have suggested
the more general
form of the DMPK equation, which reduces to  (3) in the
one-channel case\,\footnote{\,A somewhat less general
form of the equation was derived previously by Muttalib and
co-workers \cite{32,33,34} and was used in \cite{34a,34b} for
description of the conductance distribution.}
$$
\frac{\partial P\{\lambda\}}{\partial L} = \alpha\,
\sum\limits_{i}
 K_{ii} \frac{\partial }{\partial \lambda_i}
\left[\,-\gamma_i (1+2\lambda_i)\, P\{\lambda\}\right. +
$$
$$+
\left.\lambda_i (1+\lambda_i)\, J_i\{\lambda\}
\frac{\partial }{\partial \lambda_i}\,
\frac{P\{\lambda\}}{J_i\{\lambda\}} \,\right]
\eqno(25)
$$
$$
J_i\{\lambda\} = \prod\limits_{j<k}
|\lambda_j-\lambda_k|^{\beta^i_{jk}}\,,\qquad
\beta^i_{jk}=2K_{jk}/K_{ii}\,,\qquad
$$
$$
\gamma_i=(1-\sum_j K_{ij})/K_{ii} \,,\qquad
$$
where the matrix $K_{ij}$ is determined by the averaged
combinations of the $u$ and $v$ matrix elements. Equation (25)
reduces to the usual DMPK equation in the metallic regime
and provides the correct generalization beyond it. Eq.25
has the same structure for the orthogonal and unitary ensembles
and allows to describe  systems in the arbitrary magnetic field.

If the transverse size of a quasi-1D system is sufficiently
small, then its channels are well mixed by scattering, and
the approximation of equivalent channels looks reasonable,
when we can set
$\alpha K_{ii}=\tilde\alpha$, $\beta^i_{jk}=\beta$,
$\gamma_i=\gamma$. Then the evolution equation is described
by three parameters  $\tilde\alpha
L$, $\beta$, $\gamma$, which are in one-to-one
correspondence with parameters  $L/\xi$, $L_z/L$,
$L/l_H$ entering equation (10),\,\footnote{\,Analysis
of the DMPK equation shows  \cite{27} that for large number of
channels the structure of its solution does not depend on  $N$.
Formally it is obtained in the limit  $N\to\infty$, $L/a\to\infty$,
$Na/L=const$, when the DMPK equation reproduces the
diagrammatic results.} so the first problem of Sec.2 is
resolved successfully.

The second problem is also solved. Indeed, let us set
$\alpha=1/\xi_{1D}$ and consider the case $L/\xi_{1D}\ll 1$,
when the usual DMPK equation is valid; then conductance $g$
is determined by the ratio $L/\xi_{1D}$ (see (23,24))
and allows the expansion  \cite{35}
$$
g=F\left(\,\frac{L}{\xi_{1D}}\,\right)=
\frac{\xi_{1D}}{L} + B_0 +B_1 \frac{L}{\xi_{1D}} +\ldots
\eqno(26)
$$
Substituting into (14), one can easily recover the
expansion of $\beta(g)$ in the quasi-1D case
$$
\beta_{1D}(g)=-1+\frac{B_0}{g}+\frac{2 B_1}{g^2}+\ldots
\eqno(27)
$$
Accepting $\langle g \rangle$
as a typical value of $g$ and following to the Shapiro scheme
(Fig.1), one has
$$
g^{(d)}_{bL}=b^{d-1} g^{(1)}_{bL} \,.
\eqno(28)
$$
Let accept $b=1+\Delta L/L$, so that the size of the cubical
system changes from $L$ to $L+\Delta L$. Increasing the
length of the quasi-1D system is described by equation (14),
while the increase of its transverse size is
taken into account by Eq.28
$$
\ln g^{(1)}_{L+\Delta L} = \ln g^{(1)}_{L}
+\frac{\Delta L}{L} \beta_{1D}\left(g^{(1)}_{L}\right) \,,
\eqno(29)
$$
$$
\ln g^{(d)}_{L+\Delta L} = \ln g^{(1)}_{L+\Delta L}
+(d-1)\frac{\Delta L}{L} \,.
$$
As a result
$$
\ln g^{(d)}_{L+\Delta L} = \ln g^{(d)}_{L}
+\left[d-1 +\beta_{1D}\left(g^{(d)}_{L}\right) \right]
\,\frac{\Delta L}{L}
\eqno(30)
$$
and we recover the expansion (16) for
$\beta(g)=d-1+\beta_{1D}(g)$ in the
$d$-dimensional case with the coefficients
$$
A_1=B_0\,, \qquad  A_2=2B_1\,.
\eqno(31)
$$
Expansion coefficients in (26) were calculated by
Mac{$\rm\hat e$}do \cite{35} for arbitrary values of the
Wigner--Dyson parameter $\beta$:
$$
B_0=-\frac{2-\beta}{3\beta}\,,
\qquad  B_1=\frac{12-14\beta+3\beta^2}{45\beta^2}\,,
\eqno(32)
$$
which leads to
$$
A_1=-\frac{1}{3}\,,
\qquad A_2=\frac{2}{45}\,, \qquad (\beta=1)
$$
$$
A_1=0\,,
\qquad A_2=-\frac{2}{45}\,, \qquad (\beta=2)
$$
and provides the correct structure of  expansion  (16)
and results (17), (18) for the orthogonal and unitary
ensembles respectively\,\footnote{\,In the higher orders in
$\epsilon$ one should take into account
the difference of $\beta^i_{jk}$ from the
corresponding Wigner--Dyson values.}.

Let discuss once more the possibility to set $L=a$ in equation
(1), where the function  $F$ has a certain transient
behavior at scales of order  $a$. We indicated in \cite{3} that
such transient behavior can be excluded in accordance with
the general Wilson analysis  \cite{11,12}, if the special
model is chosen at small scales. Now one can see that such "ideal
model" is described by the generalized DMPK equation with
the limiting transition indicated in Footnote  4.

One can see that the Shapiro scheme becomes satisfactory in the
presence of the magnetic field, if the quasi-1D evolution
equation is taken in the form of the generalized DMPK equation.

\begin{center}
{\bf 5.  The evolution equation for $P(\rho)$} \end{center}

As was discussed in \cite{3,100},  for the correct definition
of conductance of  finite systems it is useful to introduce
semi-transparent boundaries, separating the given system from
the ideal leads attached to it. In the limit of weak
transparency one obtains universal equations, independent of the
way how the contact resistance of the reservoir is excluded
\cite{36}, which then can be
extrapolated to transparency of order unity. Such definition
surely refers to the system under consideration
(and not to the composed system "sample+ideal leads") and
provides the infinite value of conductance for ideal systems
 \cite{100}. In the limit of weakly-transparent interfaces,
 the scale of all  $\lambda_i$ increases and
the Economou--Soukoulis definition of conductance (23)
becomes equivalent to the definition
$$
g=\frac{1}{\rho} = \sum\limits_i \frac{1}{\lambda_i}\,,
\eqno(33)
$$
where conductance of each channel is taken in the Landauer
form \cite{7}.  To make transition
from $\lambda_i$  to $\rho$
let introduce the set of the "angle" variables
$\{\varphi\}=\left(\varphi_1,\varphi_2,\ldots\varphi_{N-1}\right)$
and make a change
$$
\frac{1}{\lambda_i}= \frac{1}{\rho} f_i\{\varphi\} \,.
\eqno(34)
$$
It is easy to derive for large  $\lambda_i$
$$
\rho_{ES} =\rho +\rho_0\,,\qquad \rho_0=\sum\limits_i
f_i^2\{\varphi\}   \,,
\eqno(35)
$$
and obtain the inequality
$$
0\le \rho_0\le 1\,,
\eqno(36)
$$
following from  $f_i\{\varphi\}\ge 0$ and $\sum_i
f_i\{\varphi\}=1$. In fact,
for different situations the
whole range of the $\rho_0$ values is covered. In the metallic
regime all $\lambda_i$ are of the same order, so
$f_i\{\varphi\}\sim 1/N$ and $\rho_0\sim 1/N$, which can be
arbitrary small for the large number of channels. In the strongly
localized regime conductance of the system is determined by the
single channel, so $f_1\{\varphi\}\approx 1$,
$f_i\{\varphi\}\approx 0$ ($i\ne 1$) and $\rho_0\approx 1$.

The change of variables (35) is analogous to transition
$x_i=r f_i\{\varphi\}$ from the Cartesian coordinates
$x_i$ to the
spherical ones. In the latter case the radius-vector  $r$
has a dimension of length, while the angles $\varphi_k$ are
dimensionless; correspondingly, a dimension of  $x_i$
coincides with a dimension of $r$. Analogously, it is
convenient to imagine
that $\rho$ is a dimensional quantity, while
$\lambda_i$ have a dimension of  $\rho$. Since a dimension
of each term  is conserved in transformation of derivatives,
the form of equation (25) in variables  $\rho$, $\varphi_k$ can
be written simply from dimensional considerations
$$
\frac{\partial P}{\alpha\partial L} = \,
\left[\,a_1\{\varphi\}\rho^2+a_2\{\varphi\}\rho\,\right]
\frac{\partial^2 P}{\partial \rho^2} +
$$
$$+
\left[\,a_3\{\varphi\}\rho+a_4\{\varphi\}\,\vphantom{\rho^2}\right]
\frac{\partial P}{\partial \rho} +
a_5\{\varphi\} P +
$$
$$
+\sum\limits_k
\left[\,b_k\{\varphi\}\rho+c_k\{\varphi\}\,\vphantom{\rho^2}\right]
\frac{\partial^2 P}{\partial \rho \partial \varphi_k} +
\sum\limits_k  \,g_k\{\varphi\}
\frac{\partial P}{ \partial \varphi_k} +
$$
$$+
\sum\limits_{kk'} \,h_{kk'}\{\varphi\}
\frac{\partial^2 P}{\partial \varphi_k \partial \varphi_{k'}}
\,,\eqno(38)
$$
where  terms with the factor  $1/\rho$ are
omitted\,\footnote{\,In fact, such terms should be canceled,
since there are no grounds for the term $P(\rho)\ln{\rho}$ in
the square bracket of (40).}.  Averaging over $\varphi_k$ and
eliminating the angle derivatives with the use of integration by
parts, we arrive to
$$
\frac{\partial P(\rho)}{\partial L} = \alpha\, \left[
\left(C_1\rho^2+C_2\rho\right)
\frac{\partial^2 P(\rho)}{\partial \rho^2}\right. +
$$
$$+
\left.\left(C_3\rho+C_4\vphantom{\rho^2}\right)
\frac{\partial P(\rho)}{\partial \rho}
+ C_5 P(\rho) \right]   \,.
\eqno(39)
$$
Due to conservation of probability the right hand side should have
a form of the full derivative; including  $C_1$ in redefinition
of $\alpha$, one has the equation
$$
\frac{\partial P(\rho)}{\partial L} = \tilde \alpha\,
\frac{\partial }{\partial \rho}\left[
\left(A\rho+B\vphantom{\rho^2}\right) P(\rho) +
\left(\rho^2+C\rho\right) \frac{\partial P(\rho)}{\partial
\rho} \, \right]
\eqno(40)
$$
which is of the same structure as (3). Two equations become
identical in the result of the following transformations,
carried out in  \cite{3}. Ambiguity in the
conductance definition, related with exclusion of
the reservoir contact resistance \cite{36}, corresponds
to the change  $\rho\to \rho-\rho_0$, where $\rho_0$
depends on details of a definition. This dependence is
inessential in the limit of weakly-transparent boundaries,
when a scale of $\rho$ increases unboundedly, while  $\rho_0$
remains finite. It corresponds to the following change
in the second term in the square bracket of (3)
$$
\rho(\rho+1) \,\longrightarrow
(\rho-\rho_0)(\rho+1-\rho_0) \longrightarrow\,  \rho^2 \,,
\eqno(41)
$$
leading to the universal equation, independent of $\rho_0$,
which can be extrapolated to a transparency of order
unity. Analogous procedure for the first term  in the square
bracket of (3)
$$
2\gamma\rho+\gamma \,\longrightarrow
2\gamma\rho+\gamma(1-2\rho_0) \longrightarrow\,
2\gamma\rho+\tau_0
\eqno(42)
$$
is complicated by unknown behavior of the second
summand in the course of the indicated procedure; so we denote
it as  $\tau_0$, bewaring of setting to be zero. The further
analysis shows that $\tau_0$ should be  considered as finite on
the physical grounds. As a result, the following
modification of equation (3) is practically used in the Shapiro
scheme
$$
\frac{\partial P(\rho)}{\partial L} = \tilde\alpha\,
\frac{\partial }{\partial \rho}
\left[\,-(2\gamma\rho+\tau_0) P(\rho) + \rho^2
\frac{\partial }{\partial \rho} P(\rho) \,\right].
\eqno(43)
$$
The analogous procedure applied to Eq.40 leads to the same
result, if we observe that the  parameter $C$ is restricted by
the inequality $0\le C\le 1$ (see Appendix 2) and cannot leads
to unpredictable effects.

In the previous paper  \cite{3} we did not consider $\tau_0$
as an essential parameter.  It arises in the course of the
ill-defined extrapolation procedure  to transparency of order
unity and specifies the absolute scale of conductance, which is
not controlled in theory. The situation is changed in the
presence of the magnetic field, since $\tau_0$ may depend on the
magnitude of the field and should be considered as an essential
parameter. Therefore, the evolution equation for
$P(\rho)$ is determined by three parameters   $\tilde\alpha L$,
$\gamma$, $\tau_0$, which are in one-to-one correspondence
with parameters  $L/\xi$, $L_z/L$, $L/l_H$ entering (10).

As a result, the first problem of Sec.2 is solved in
the framework of equation for $P(\rho)$, but the second problem
still exists.  Indeed, it is easy to verify that transformation
of type (13) does not  allow
the result $\nu=1/2\epsilon$  for any constant values of
$A$, $B$, $C$ in (40). However, these
coefficients can be considered as constant only in the case when
distribution $P(\rho)$ is almost stationary, and the
$\varphi_k$ distribution has a time to relax. In the quasi-1D
geometry a stationary distribution $P(\rho)$ is never
realized, and coefficients $A$, $B$, $C$ always contain a
certain dependence on $L$. If such dependence is taken into
account, there is no problem to reach the result
$\nu=1/2\epsilon$ (see Appendix 3).

We can conclude, that two problems of Sec.2 can be solved
on the level of the  $P(\rho)$  evolution equation, if
some of statements made in  \cite{3} are formulated more
accurately.  However, it requires the certain assumptions,
which
fulfil automatically on the level of the DMPK equation.

\begin{center}
{\bf 6. Transition to the $d$-dimensional case}
\end{center}

If the quasi-1D evolution equation is accepted in the form
(43), then the Shapiro scheme (Fig.1) allows  a simple transition
to the  $d$-dimensional case. The equation for $W(g)$,
corresponding to (43), is obtained by replacements
$P=g^2W$, $\rho=1/g$ \cite{3}
$$
\frac{\partial W(g)}{\partial L} =\tilde\alpha
\left[\,\left(2\gamma g+2g +\tau_0 g^2\right) W(g) +g^2
W_g'(g)\,\right]'_g \,,
\eqno(44)
$$
and to solve the $d$-dimensional problem, one should find
a distribution of the sum of $n=b^{d-1}$ independent
random quantities with the same distribution $W(g)$: it is made
by introducing the characteristic function  $F(t)=\left\langle
e^{igt} \right\rangle$ and raising it to the power  $n$.
Instead of the characteristic function it is more convenient
to use the Laplace transform
$$
F(\tau) = \int_0^\infty e^{-\tau g} W(g) dg \,,
$$
while the indicated procedure can be realized in the
differential form. Equation for  $F(\tau)$ corresponding
to (44) can be written for finite differences
$$
F_{L+\Delta L}(\tau) = F_{L}(\tau) + \tilde\alpha \Delta L
\left[\,\tau(\tau+\tau_0) F''_{L}(\tau)\right. -
$$
$$-\left.
2\gamma \tau F'_{L}(\tau) \right] \,.
\eqno(45)
$$
Raising $F_{L}(\tau)$ to the power $n=b^{d-1}$ and setting
$b=1+\Delta L/L$, we obtain the additional term
$(\Delta L/L)(d\!-\!1) F_L \ln F_L$ in the right hand side of
(45). As a result, the evolution equation for
the $d$-dimensional system has a following form
$$
\frac{\partial F(\tau)}{\partial \ln L} = \tilde\alpha L
\left[\,\tau(\tau+\tau_0) F''(\tau) -
2\gamma \tau F'(\tau)\right. +
$$
$$+\left.
p F(\tau) \ln F(\tau) \right] \,,
\eqno(46)
$$
where  $p=(d\!-\!1)/\tilde\alpha L$. The quantity
$\tilde\alpha L$ has a sense of $L/\xi$ and evolution in  $L$
for fixed $L/\xi$ leads to a stationary distribution,
corresponding to large scales; in the course of this evolution
parameters $\tau_0$, $\gamma$ and $p$ tend to constant
limits  and variability of the coefficients in
the quasi-1D equation (40) is of no significance. Eq.46
describes the transient process due to increasing of   $L$ from
the atomic scale to the scales of order $\xi$, and its
stationary version is of the main interest; as a result, the
parameter $\tilde\alpha L$ looses its actuality and its
role comes to the parameter $p$. Therefore, the stationary values
of  $\tau_0$, $\gamma$, $p$ are in one-to-one correspondence with
parameters $L/\xi$, $L_z/L$, $L/l_H$ of Eq.10, which are
maintained fixed in the limiting transition $L\to\infty$.
As a result,  $\xi$ increases unboundedly and all
obtained distributions refer to the critical point.
"The critical distribution" discussed usually in
theoretical papers  (e.g. \cite{4,5}) corresponds to
the values $L/\xi=0$, $L_z/L=1$, $L/l_H=\infty$
(in the finite magnetic field) or $L/l_H=0$ (in the absence
of a field), which fix the critical values of parameters
$\tau_0$, $\gamma$, $p$ for the corresponding dimension of space.
For large but finite systems, any values of  $L/\xi$, $L_z/L$,
$L/l_H$ are accessible.

The stationary version of equation (46) was extensively
studied in the paper \cite{3}.   The universal property of
distributions is existence of two asymptotic regimes, log-normal
for small $g$ and exponential for large $g$, while their
actuality depends on the specific situation. In the metallic
phase, a distribution is determined by the central Gaussian peak,
while two asymptotic regimes refer to its far tails. In the
critical region, the log-normal behavior is extended to a
vicinity of the maximum,  and
practically all distribution is determined by two asymptotes. In
proceeding to the localized phase, the log-normal behavior
extends even more and forces out the exponential asymptotics to
the region of the remote tail.

\begin{center}
{\bf 7. Conclusion}
\end{center}

It should be clear from preceding, that the
conductance distribution in the magnetic field is described by
the same equations as in  absence of a field. Variation  of
the magnetic field does not lead to any qualitative effects in
the conductance distribution and only changes its quantitative
characteristics, moving the system position in the
three-parameter space of $\tau_0$, $\gamma$, $p$.  This
conclusion is in accordance with numerical experiments
\cite{9,10}. The change of the distribution under variation of
the field is well-known for the metallic phase: the magnetic field
increases the mean value of conductance (negative
magneto-resistance \cite{37}) and diminishes its root-mean-square
fluctuation \cite{1,2,38} due to suppression of the Cooperon
contributions.

In contrast to the previous paper \cite{3}, the quasi-1D
evolution equations were established on the basis of the
generalized DMPK equation, and not by a simple analogy with
one-dimensional systems. It gave possibility to refine some
statements of \cite{3} and formulate them more accurately.
In the spirit of \cite{3}, we did not try to estimate the
parameters $\tau_0$, $\gamma$, $p$ for any specific
situations, but studied all family of distributions in whole.
The values of these parameters can be established by
calculation of several  moments of conductance,
which is possible by the standard methods.

\begin{center}
{\it Appendix 1.} One-dimensional evolution equation
\end{center}

To establish the general form of the transfer matrix, we note
that amplitudes of the incident and transmitted waves are related
by the scattering  $S$-matrix
$$
\left ( \begin{array}{cc} B \\ C \end{array} \right)\,
=  S \left ( \begin{array}{cc} A \\ D \end{array}
\right)
=  \left ( \begin{array}{cc} r & t' \\
t & r' \end{array} \right)\,
\left ( \begin{array}{cc} A \\ D \end{array}
\right) \,,
\eqno(A.1)
$$
which is defined by the amplitudes of transmission   ($t$) and
reflection ($r$) for waves incident from the left of scatterer,
and analogous amplitudes  ($t'$ and $r'$) for waves incident
from the right. The unitarity of $ S$-matrix gives
$$
|r|^2+|t|^2=1\,,\quad  |r'|^2+|t'|^2=1\,,\quad
r^*t' = -t^* r' \,.
\eqno(A.2)
$$
Squaring the modulus of the latter relation, one has
$|r|=|r'|$, $|t|=|t'|$ and the elements of
$S$-matrix can be written as
$$
r=|r|e^{i\theta},\quad
r'=|r|e^{i\theta'},\quad
t=|t|e^{i\varphi},\quad
t'=|t|e^{i\varphi'},
\eqno(A.3)
$$
with the additional relation for phases
$$
e^{i\theta+i\theta'-i\varphi-i\varphi'} =-1\,,
\eqno(A.4)
$$
following from $(A.2)$. Rewriting $(A.1)$ in the form (4),
we have
$$
\hat T= \left ( \begin{array}{cc} 1/t & - r'/t \\
 r/t & (tt'-rr')/t \end{array} \right)\,
=
$$
$$
=\left ( \begin{array}{cc} |1/t| e^{-i\varphi} &
|r/t|  \,e^{-i\theta+i\varphi'}
\\ |r/t|  \,e^{i\theta-i\varphi}
 & |1/t|  \,e^{i\varphi'}
 \end{array} \right)\,.
\eqno(A.5)
$$
Introducing the Landauer resistance $\rho=|r/t|^2$ \cite{7},
setting $\zeta=\varphi'-\varphi$ and shifting the origin of
$\theta$, one can reduce $(A.5)$ to the form (8).

The time-reversal symmetry is in fact the invariance
with respect to
the complex conjugation, where $S$ transforms to $S^*$ and the
incident and reflected waves change their places. It gives the
relation $S^*=S^{-1}$, which leads to $S=S^{T}$ and
$\varphi=\varphi'$, if the unitary condition $S^+=S^{-1}$  is
taken into account. The analogous, but  more complicated
analysis allows to establish the canonical representation (22)
with the additional relations $u=v^*$, $u_1=v_1^*$  in the
time-reversal case.

If a length $L$ of an one-dimensional system is increased to
$L+\Delta L$, then the transfer matrices are multiplied, $\hat
T_{L+\Delta L}=\hat T_{L} \hat T_{\Delta L}$. Accepting the form
(8) for the matrix $\hat T_L$ and setting$$
\hat T_{\Delta L}=
 \left ( \begin{array}{cc} \sqrt{1\!+\!\epsilon^2}\, e^{i\beta_1}
 & -i\epsilon \,e^{i\beta_2+i\beta_3} \\ i\epsilon\,
 e^{-i\beta_2} & \sqrt{1\!+\!\epsilon^2}\, e^{-i\beta_1+i\beta_3}
 \end{array} \right)\,,
\eqno(A.6)
$$
where $\epsilon$, $\beta_1$, $\beta_2$, $\beta_3$ are small
random quantities,\,\footnote{\,The form of
the matrix $(A.6)$ is chosen from the analogy with a point
scatterer, allowing to accept a zero value for the
mean of $\epsilon$  \cite{3}.}
we have for the parameter $\tilde\rho$,
corresponding to $\hat T_{L+\Delta L}$, in the second order in
$\epsilon$
$$
\tilde \rho=\rho-2\epsilon\sqrt{\rho(\rho+1)}
\sin\psi +\epsilon^2 (2\rho+1)\,,
\eqno(A.7)
$$
with
$$
\psi=\theta-\varphi-\zeta+\beta_1+\beta_2\,.
\eqno(A.8)
$$
Relation $(A.7)$ differs from $(A.2)$  of
\cite{3} only by definition of the phase $\psi$, and the
rest of derivation  coincides
with that in \cite{3}, leading to equation  (3)
with parameters  (9).

\begin{center}
{\it Appendix 2.} Inequality for the parameter $C$ in (40)
\end{center}

If $\varphi_{1}$, $\varphi_{2}$, $\ldots$, $\varphi_{N-1}$
are independent variables, then setting
$$
\frac{1}{\lambda_i}\,= \,\frac{\varphi_i}{\rho}\,,\quad
i=1,2,\ldots,N-1   \qquad\qquad\qquad\qquad
$$
$$
\frac{1}{\lambda_N}\,= \,\frac{\varphi_N}{\rho}\,,\qquad
\varphi_N=1-\varphi_1-\varphi_2-\ldots-\varphi_{N-1}
 \eqno(A.9)
$$
and coming from  $\lambda_i$ to  $\rho$, $\varphi_1$,
$\varphi_2$, $\ldots$, $\varphi_{N-1}$, we obtain the
following form for the first term in the right hand side of  (38)
$$
\left[\,\rho^2\sum\limits_{i=1}^N K_{ii} \varphi_i^2 +
\rho\sum\limits_{i=1}^N K_{ii} \varphi_i^3\,\right]
\frac{\partial^2 P}{\partial \rho^2}
\,.\eqno(A.10)
$$
Since $K_{ii}\ge 0$ (see Eq.34 in  \cite{8}) and
$0\le \varphi_i\le 1$, then
$$
\sum\limits_{i=1}^N K_{ii} \varphi_i^2 \, \ge  \,
\sum\limits_{i=1}^N K_{ii} \varphi_i^3 \,\ge \, 0
\,,\eqno(A.11)
$$
and averaging over $\varphi_k$ gives $C_1\ge C_2 \ge 0$
in (39) and  $0\le C\le 1$  in (40).

\vspace{5mm}

\begin{center}
{\it Appendix 3.} To estimation of the critical behavior
 \end{center}

Multiplying (40) by $\rho$ and integrating, one has the closed
equation for $\langle \rho\rangle$
$$
\frac{\partial \langle \rho\rangle}{\partial L} =
a\langle \rho\rangle +b
\eqno(A.12)
$$
with $a=\tilde\alpha(2-A)$, $b=\tilde\alpha(C-B)$.
Setting
$$
a=a_0+a_1 L\,,\qquad b=b_0+b_1 L  \,,
\eqno(A.13)
$$
we have
$$
 \langle \rho\rangle = b_0 L +
 {\textstyle \frac{1}{2}} (a_0 b_0+b_1) L^2+
 {\textstyle \frac{1}{6}} (a_0^2 b_0+a_0 b_1+2a_1 b_0) L^3+\ldots
\eqno(A.14)
$$
and the choice  $a_0 b_0+b_1=0$ eliminates the term of order
$L^2$. For a narrow distribution it is equivalent to
disappearance of $B_0$ in (26) and validity of the result (18).

\end{document}